\documentclass[utf8]{FrontiersinVancouver} 
\usepackage{url,hyperref,lineno,microtype}
\usepackage{subcaption}
\usepackage[onehalfspacing]{setspace}
\usepackage{comment}
\usepackage{xcolor}
\usepackage{lipsum}
\usepackage{todonotes}
\usepackage{multirow}
\usepackage{subfig}
\usepackage{graphicx}
\usepackage{booktabs}
\usepackage{threeparttable}
\usepackage{pifont}
\newcommand{\mycmark}{\ding{51}}
\newcommand{\myxmark}{\ding{55}}
\usepackage{eurosym}
\usepackage{tikz}
\usetikzlibrary{calc}
\usepackage{pgfplots}
\pgfplotsset{compat=1.18}
\usepgfplotslibrary{dateplot}
\usepackage{siunitx}
\DeclareSIUnit\flop{FLOP}
\DeclareSIUnit[per-mode=symbol]\floppersec{\flop\per\second}
\ifx\myfixmestatus\undefined
\usepackage[status=draft]{fixme}
\else
\usepackage[status=\myfixmestatus]{fixme}
\fi

\fxsetup{%
  multiuser,
  layout=inline,
  theme=color,
  inlineface=\bfseries,
  envface=\bfseries
}
\FXRegisterAuthor{es}{aes}{ES} % Estela Suarez (JSC)
\FXRegisterAuthor{mf}{amf}{MF} % Martin Frank (KIT)
\FXRegisterAuthor{ne}{ane}{NE} % Norbert Eicker (JSC)
\FXRegisterAuthor{bv}{abv}{BV} % Benedikt von St. Vieth (JSC)
\FXRegisterAuthor{ah}{aah}{AH} % Andreas Herten (JSC)
\FXRegisterAuthor{kt}{akt}{KT} % Kay Thust (JSC)
\FXRegisterAuthor{pf}{apf}{PF} % Peter Frech (JSC)
\FXRegisterAuthor{tf}{atf}{TF} % Tom Fieseler (JSC)
\FXRegisterAuthor{cm}{acm}{CM} % Cristina Manzano (JSC)
\FXRegisterAuthor{ss}{ass}{SS} % Salem El Sayed (JSC)
\FXRegisterAuthor{mo}{amo}{MO} % Michael Ott (LRZ)
\FXRegisterAuthor{ti}{ati}{TI} % Thomas Ilsche (TUD)
\FXRegisterAuthor{so}{aso}{SO} % Sebastian Oeste (TUD)
\FXRegisterAuthor{dh}{adh}{DH} % Daniel Hackenberg (TUD)
\FXRegisterAuthor{gh}{agh}{GH} % Georg Hager (FAU)
\FXRegisterAuthor{je}{aje}{JE} % Jan Eitzinger (FAU)
\FXRegisterAuthor{bk}{abk}{BK} % Bastian Köller (HLRS)
\FXRegisterAuthor{rs}{ars}{RS} % Ralf Schneider (HLRS)
\FXRegisterAuthor{hb}{ahb}{HB} % Hendryk Bockelmann (DKRZ)
\FXRegisterAuthor{pg}{apg}{PG} % Pay Giesselmann (DKRZ)
\FXRegisterAuthor{el}{ael}{EL} % Erwin Laure (MPCDF)
\FXRegisterAuthor{kr}{akr}{KR} % Klaus Reuter (MPCDF)

\def\keyFont{\fontsize{8}{11}\helveticabold }
\def\firstAuthorLast{Suarez {et~al.}} %use et al only if is more than 1 author
\def\Authors{Estela Suarez\,$^{1,2,3*}$, Hendryk Bockelmann\,$^{4}$, Norbert Eicker\,$^{1,5}$, Jan Eitzinger\,$^{6}$, Salem El Sayed\,$^{1}$, Thomas Fieseler\,$^{1}$, Martin Frank\,$^{7}$, Peter Frech\,$^{1}$, Pay Giesselmann\,$^{4}$, Daniel Hackenberg\,$^{8}$, Georg Hager\,$^{6}$, Andreas Herten\,$^{1}$, Thomas Ilsche\,$^{8}$, Bastian Koller\,$^{9}$, Erwin Laure\,$^{10}$, Cristina Manzano\,$^{1}$, Sebastian Oeste\,$^{8}$, Michael Ott\,$^{11}$, Klaus Reuter\,$^{10}$, Ralf Schneider\,$^{9}$, Kay Thust\,$^{1}$, Benedikt {von St. Vieth}\,$^{1}$
} 

\def\Address{$^{1}$J\"{u}lich Supercomputing Centre~(JSC), Forschungszentrum J\"{u}lich GmbH, J\"{u}lich, Germany \\
$^{2}$SiPEARL GmbH, Duisburg, Germany  \\
$^{3}$Institute of Computer Science, University of Bonn, Bonn, Germany\\
$^{4}$Deutsches~Klimarechenzentrum~GmbH~(DKRZ), Hamburg, Germany \\
$^{5}$Bergische Universt\"{a}t Wuppertal, Wuppertal, Germany \\
$^{6}$Erlangen National High Performance Computing Center~(NHR@FAU), Erlangen, Germany \\
$^{7}$Karlsruhe Institute of Technology~(KIT), Karlsruhe, Germany \\
$^{8}$ZIH, CIDS, TU Dresden, Dresden, Germany \\
$^{9}$High-Performance Computing Center Stuttgart (HLRS), Stuttgart, Germany \\
$^{10}$Max Planck Computing and Data Facility (MPCDF), Garching, Germany \\
$^{11}$Leibniz Supercomputing Centre (LRZ), Garching, Germany
}
\def\corrAuthor{Estela Suarez, J\"{u}lich Supercomputing Centre, Forschungszentrum J\"{u}lich GmbH, Leo Brandt Str, 52428 J\"{u}lich, Germany}

\def\corrEmail{e.suarez@fz-juelich.de}

\begin{document}
\onecolumn
\firstpage{1}

\title[Energy efficient HPC in Germany]{Energy-aware operation of HPC systems in Germany} 

\author[\firstAuthorLast ]{\Authors} %This field will be automatically populated
\address{} %This field will be automatically populated
\correspondance{} %This field will be automatically populated

\extraAuth{}% If there are more than 1 corresponding author, comment this line and uncomment the next one.

\maketitle
\begin{abstract}
High-Performance Computing (HPC) systems are among the most energy-intensive scientific
facilities, with electric power consumption reaching and often exceeding 20 megawatts per installation.
Unlike other major scientific infrastructures such as particle accelerators or high-intensity light sources, which are few around the world, the number and size of supercomputers are continuously increasing. Even if every new system generation is more energy efficient than the previous one, the overall growth in size of the HPC infrastructure, driven by a rising demand for computational capacity across all scientific disciplines, and especially by artificial intelligence workloads (AI), rapidly drives up the energy demand. This challenge is particularly significant for HPC centers in Germany, where high electricity costs, stringent national energy policies, and a strong commitment to environmental sustainability are key factors. This paper describes various state-of-the-art strategies and innovations employed to enhance the energy efficiency of HPC systems within the national context. Case studies from leading German HPC facilities illustrate the implementation of novel heterogeneous hardware architectures, advanced monitoring infrastructures, high-temperature cooling solutions, energy-aware scheduling, and dynamic power management, among other optimizations. 
By reviewing best practices and ongoing research, this paper aims to share valuable insight with the global HPC community, motivating the pursuit of more sustainable and energy-efficient HPC operations.

\tiny
 \keyFont{ \section{Keywords:} High-Performance Computing, HPC, energy efficiency, data centre, cooling, monitoring, hardware, heterogeneous compute architectures, software, Germany} %All article types: you may provide up to 8 keywords; at least 5 are mandatory.
\end{abstract}

\section{Introduction} % - [500 words]}
\label{sec:intro}

High-Performance Computing (HPC) systems are indispensable instruments in scientific research, but at the same time  energy-hungry infrastructures. Although the computational capacity per Watt of computers and processing units is improving over time~\citep{green500, koomey2011}, the demand for more compute capacity --~recently strongly driven by large-scale computations for Artificial Intelligence (AI)~-- is growing disproportionate, leading to the deployment of even more computing infrastructures and services (e.g.~\citep{AIfact}). Existing and upcoming HPC facilities must provide significant computational power and, consequently, require large amounts of energy for IT and cooling, making their sustainability a major concern.

The two most significant issues pertaining to these energy demands are the economic viability of maintaining the necessary infrastructure in the light of rising electricity prices, and the considerable environmental impact of generating this electricity, which raise concerns in society. Germany is characterised by higher-than-average electricity costs in comparison to other countries, which is attributed in part to its particular energy mix and reliance on imported resources. To address societal concerns, policies have been implemented with the objective of addressing environmental impact. The German Energy Efficiency Act~\citep{eneffg} mandates that businesses, in particular commercial datacenters, and research institutions, including supercomputing centres, observe stricter energy consumption limits. Furthermore, the European Supply Chain Directive~\citep{EUsuplchain} has the objective of accounting for the energy consumed and CO$_2$ generated during the entire life cycle of a given product, from the moment of fabrication until its end of life. For hosting sites to estimate the \textit{embedded carbon footprint} in HPC systems, they must rely on manufacturers to explicitly state the embedded carbon in their products. From all the above it becomes imperative that energy efficiency measures have to be integrated into the design and operation of HPC sites in order to align with national and EU-wide sustainability goals and legislation. 

This paper examines the latest developments and trends in energy efficiency at major German HPC centres, including DKRZ (Deutsches Klimarechenzentrum GmbH), FAU (Friedrich-Alexander-Universität Erlangen-Nürnberg), HLRS (High-Performance Computing Centre Stuttgart), JSC (J\"{u}lich Supercomputing Centre), KIT (Karlsruhe Institute of Technology), LRZ (Leibniz Supercomputing Centre), MPCDF (Max Planck Computing and Data Facility) and TUD (Technische Universität Dresden, Center for Interdisciplinary Digital Sciences (CIDS), Center for Information Services and High Performance Computing (ZIH)).

These institutions host and operate supercomputing facilities at European, German, and regional level, with 20 machines ranging from ranking 21 (JUWELS Booster at JSC) to 446 (HoreKa-Blue at KIT) in the TOP500 list of June 2024~\citep{top500}. All these HPC centres balance the need for advanced computing infrastructure with sustainability efforts by exploring cutting-edge solutions such as energy-efficient hardware, advanced system monitoring and management, and optimised cooling systems. By presenting the experience applied by these institutions in production environments, we aim at inspiring others on applying similar techniques to reduce the energy footprint of HPC infrastructures.

The contributions of this paper are:
\begin{itemize}
    \item Report on the energy consumption and cost  trends in the deployment and operation of HPC systems.
    \item Describe energy efficiency trends at the infrastructure, system, monitoring, and software level.
    \item Give an overview of energy efficiency approaches at a number of major German HPC centres, covering European, national, and regional facilities.
    \item Report on current techniques applied already in production, approaches explored in research activities, and observed trends.
\end{itemize}

\section{Motivation and Cost trends} % - [\textbf{1000 words}]}
\label{sec:motiv}

Over the past decade, computing performance of HPC systems has continued to grow exponentially, driven by advances in microarchitecture and semiconductor technology~\citep{top500}. Manufacturing processes have been scaled down to single-digit nanometer units~\citep{tsmc}. Finer structures allow more transistors per chip, increasing computing power per physical area and per watt, since signals have to travel shorter distances across the chip. However, with the breakdown of Dennard scaling and Moore’s Law also coming to an end, the power draw of HPC systems is steadily increasing. First Exascale systems exceed \qty{20}{\mega\watt}, and further gain in compute performance in newer machines is mostly driven by larger system sizes. This comes at the prize of higher power consumption and poses challenges for HPC operations at scale in terms of sustainability, economic viability, and technical feasibility. 

In particular, the increasing performance has led to an increase in power density, making traditional air cooling methods inadequate for heat dissipation. Direct liquid cooling, particularly with hot water, is advantageous in terms of energy reuse and infrastructure efficiency (see section~\ref{sec:cool}). However, as components such as GPUs generate more and more heat, the demand for chilled water cooling systems has increased, creating a mismatch between cooling technology trends and energy efficiency goals.

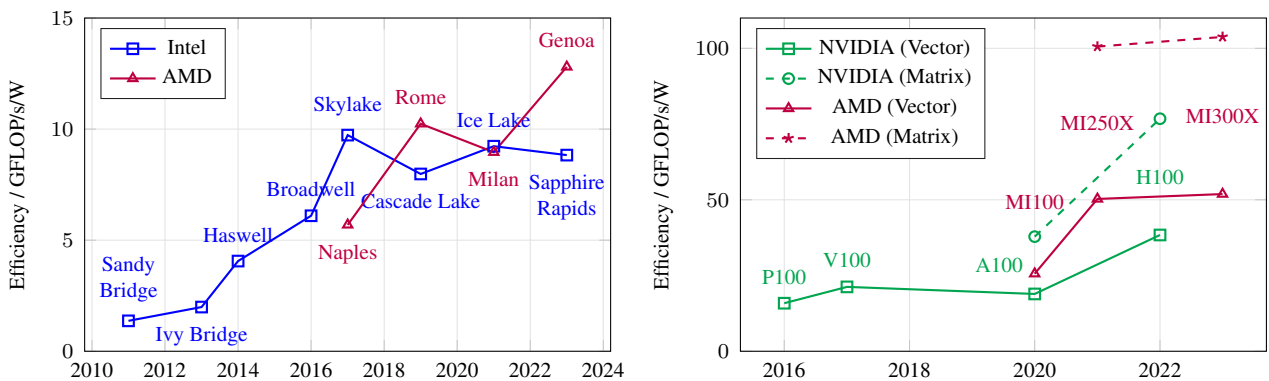
\begin{figure}[!h]
    \colorlet{colnvidia}{Green}%
    \colorlet{colamd}{purple}%
    \colorlet{colintel}{blue}%
    \centering%
        \begin{tikzpicture}
        \begin{axis}[
            width=0.48\linewidth, % Width of the plot
            height=6cm, % Height of the plot
            ylabel={Efficiency / GFLOP/s/W},
            date coordinates in=x,
            %xmin=2016, xmax=2023, % X-axis range
            ymin=0, ymax=15, % Y-axis range, modify based on your actual data
            xtick={2010-01-01, 2012-01-01, 2014-01-01, 2016-01-01, 2018-01-01, 2020-01-01, 2022-01-01, 2024-01-01}, % Define years as date coordinates 
            %xticklabels={2016, 2018, 2020, 2022}, % Show only the year in the label
            legend pos=north west,
            grid=both,
            xticklabel={\year},
            major grid style={line width=.2pt,draw=gray!50},
            minor grid style={line width=.1pt,draw=gray!30},
            label style={font=\footnotesize},
            tick label style={font=\footnotesize},
            legend style={%
                font=\footnotesize
            },
        ]

        % Product Line 1 Data
        \addplot[color=colintel, mark=square, thick]
        coordinates {
            (2011-01-01,1.37)
            (2013-01-01,1.99)
            (2014-01-01,4.06)
            (2016-01-01,6.10)
            (2017-01-01,9.73)
            (2019-01-01,7.98)
            (2021-01-01,9.23)
            (2023-01-01,8.83)
        } 
        node [above=1mm, font=\footnotesize, align=center] at (axis cs:2011-01-01,1.37) {Sandy\\Bridge}
        node [below=1mm, font=\footnotesize] at (axis cs:2013-01-01,1.99) {Ivy Bridge}
        node [above=1mm, font=\footnotesize] at (axis cs:2014-01-01,4.06) {Haswell}
        node [above=1mm, font=\footnotesize] at (axis cs:2016-01-01,6.10) {Broadwell}
        node [above=1mm, font=\footnotesize] at (axis cs:2017-01-01,9.73) {Skylake}
        node [below=1mm, font=\footnotesize] at (axis cs:2019-01-01,7.98) {Cascade Lake}
        node [above=1mm, font=\footnotesize] at (axis cs:2021-01-01,9.23) {Ice Lake}
        node [below=1mm, font=\footnotesize, align=center] at (axis cs:2023-01-01,8.83) {Sapphire\\Rapids}
        ;
        \addlegendentry{Intel}
        \addplot[color=colamd, mark=triangle, thick]
        coordinates {
          (2017-01-01,5.69)
          (2019-01-01,10.24)
          (2021-01-01,8.96)
          (2023-01-01,12.80)
        }
        node [below=1mm, font=\footnotesize] at (axis cs:2017-01-01,5.69) {Naples}
        node [above=1mm, font=\footnotesize] at (axis cs:2019-01-01,10.24) {Rome}
        node [below=1mm, font=\footnotesize] at (axis cs:2021-01-01,8.96) {Milan}
        node [above=1mm, font=\footnotesize] at (axis cs:2023-01-01,12.80) {Genoa}
        ;
        \addlegendentry{AMD}
        \end{axis}
    \end{tikzpicture}
    \begin{tikzpicture}
        \begin{axis}[
            width=0.48\linewidth, % Width of the plot
            height=6cm, % Height of the plot
            ylabel={Efficiency / GFLOP/s/W},
            date coordinates in=x,
            %xmin=2016, xmax=2023, % X-axis range
            ymin=0, ymax=110, % Y-axis range, modify based on your actual data
            xtick={2016-01-01,2018-01-01,2020-01-01,2022-01-01}, % Define years as date coordinates 
            %xticklabels={2016, 2018, 2020, 2022}, % Show only the year in the label
            legend pos=north west,
            grid=both,
            xticklabel={\year},
            major grid style={line width=.2pt,draw=gray!50},
            minor grid style={line width=.1pt,draw=gray!30},
            label style={font=\footnotesize},
            tick label style={font=\footnotesize},
            legend style={%
                font=\footnotesize
            }
        ]
        \addplot[color=colnvidia, mark=square, thick]
        coordinates {
             (2016-01-01, 15.86)
             (2017-01-01, 21.25)
             (2020-01-01, 18.92)
             (2022-01-01, 38.38)
        } 
        node [above=1mm, font=\footnotesize] at (axis cs:2016-01-01,15.86) {P100}
        %node [coordinate, pin={[pin distance=2mm]above:{V100}}] at (axis cs:2017-01-01,27.26) {}
        node [above=1mm, font=\footnotesize] at (axis cs:2017-01-01,21.25) {V100}
        node [left, font=\footnotesize] at ($(axis cs:2020-01-01,18.92)!0.5!(axis cs:2020-01-01,37.84)$) {A100}
        node [left, font=\footnotesize, anchor=center] at ($(axis cs:2022-01-01,38.38)!0.5!(axis cs:2022-01-01,76.76)$) {H100};
        \addlegendentry{NVIDIA (Vector)}
        \addplot[color=colnvidia, mark=o, dashed, thick, mark options={solid}]
        coordinates {
             (2020-01-01, 37.84)
             (2022-01-01, 76.76)
        };
        \addlegendentry{NVIDIA (Matrix)}
        \addplot[color=colamd, mark=triangle, thick] coordinates {
            (2020-01-01, 25.6)
            (2021-01-01, 50.28)
            (2023-01-01, 51.89)
        }
        node [above=7mm, font=\footnotesize] at (axis cs:2020-01-01,25.6) {MI100}
        node [font=\footnotesize, anchor=center] at ($(axis cs:2021-01-01,50.28)!0.5!(axis cs:2021-01-01,100.57)$) {MI250X}
        node [font=\footnotesize, anchor=center] at ($(axis cs:2023-01-01,51.89)!0.5!(axis cs:2023-01-01,103.76)$) {MI300X}
        ;
        \addlegendentry{AMD (Vector)}
        \addplot[color=colamd,mark=star,dashed,thick]
        coordinates {
            (2021-01-01, 100.57)
            (2023-01-01, 103.76)
        };
        \addlegendentry{AMD (Matrix)}
        \end{axis}
    \end{tikzpicture}
    \caption{Energy efficiency for FP64 floating point throughput of a selection of CPUs (left) and GPUs (right). Determined with theoretical peak performance and TDP of one socket/GPU using the highest SKU of each generation. For CPUs, the frequency used to determine peak performance is the lowest frequency measured with a very hot benchmark. For GPUs, the base frequency is taken, assuming continued computations. For GPUs results with and without considering tensor cores are shown.  The graphs compare similar, albeit not identical frequency types (measured vs. computed); cross-graph comparability is only limited.}
    \label{fig:EnergyEfficiencyChips}
\end{figure}

Figure \ref{fig:EnergyEfficiencyChips} shows the development of energy efficiency in terms of double precision (DP) GFLOP/s/Watt over the last decade for CPUs and GPUs.
Intel was able to deliver a steady increase in energy efficiency up to the Skylake micro-architecture in 2017.
Still this development stagnated in the last five years, because Intel struggled to further improve and shrink their in-house manufacturing process.
AMD on the other hand, due to technological limitations, was forced to go for a multi-die setup, which turned out to be the best choice as it allowed them to quicker employ more energy efficient manufacturing processes, while at the same time delivering higher core counts, which Intel was not able to deliver with their monolithic chip designs.
One reason for this was that the Intel core is more complex and requires a larger die area than the AMD counterpart.
2025 will show if multi-core chips can provide significant energy improvements again.
Intel has addressed their shortcomings: They developed a smaller core, also employ a multi-chip setup, and again are catching up with a competitive in-house manufacturing process.
AMD for the first time employs full-width AVX512 execution units, and it has to be seen if they can implement them in an energy-efficient manner without sacrificing frequency.
Independent of what efficiency improvements multi-core CPUs can deliver, Figure \ref{fig:EnergyEfficiencyChips} also shows that they will never be competitive against more specialized GPU accelerators.
It is interesting that while GPUs deliver a superior energy efficiency compared to multi-core CPUs, the efficiency did not improve significantly over the last 10 years.
Both GPU vendors provide even higher efficiencies when using the more specialized tensor core units.
This leads to the conclusion that the largest increases in energy efficiency are enabled by specialization.

\begin{figure}[!h]
\centering
\begin{tikzpicture}[]
    \definecolor{clr1}{HTML}{007DCD}
    \definecolor{clr2}{HTML}{92D050}
    \definecolor{clr3}{HTML}{49721E}
    \definecolor{clr4}{HTML}{FF9933}
	%\pgfplotsset{cycle list/RdYlBu-4}
	\begin{axis}[
        height=5cm,
        scale only axis=true,
		ybar stacked,
		x=30pt,
		%ybar=5pt,
		%bar width=15pt,
		%nodes near coords,
		%enlargelimits=0.15,
		legend pos=north west,
		legend cell align={left},
		legend style={%
		%   at={(0.5,-0.20)},
		%   anchor=north,%
		%   legend columns=-1,
		%legend columns=4,
		%at={(xticklabel cs:0.5)},
		%anchor=north,
		%draw=none,
		%fill=none
		},
		ymajorgrids,
		tick pos=lower,
		grid style={draw=gray!50},
		ymin=0,
		ylabel={Average Electricity Price / €/MWh},
		ylabel style={font=\small},
		symbolic x coords={2014, 2015, 2016, 2017, 2018, 2019, 2020, 2021, 2022, 2023, 2024},
		xtick=data,
		ytick={0, 100, 200, 300, 400},
		%x tick label style={rotate=45,anchor=east}, 
		legend style={font=\small, draw=none},
		ticklabel style = {font=\small},
		%cycle list name=colorbrewer,
		legend image code/.code={%
			\draw[#1] (0cm,-0.1cm) rectangle (0.3cm,0.1cm);
		},
		every axis plot/.append style={fill},
		%cycle list name=RdYlBu-4,
        y axis line style={draw opacity=0}
	]
    \pgfplotsset{cycle list shift=1}
	\addplot+[ybar, fill=clr1, draw=clr1] plot coordinates {(2014,69.5) (2015,71.9) (2016,70) (2017,80.2) (2018,89.7) (2019,94.8) (2020,84.8) (2021,123) (2022,386.2) (2023,216) (2024,151.6)};
	\addplot+[ybar, fill=clr2, draw=clr2] plot coordinates {(2014,62.4) (2015,61.7) (2016,63.54) (2017,68.8) (2018,67.92) (2019,64.05) (2020,67.56) (2021,65) (2022,18.62) (2023,0) (2024,0)};
	\addplot+[ybar, fill=clr3, draw=clr3] plot coordinates {(2014,5.89) (2015,3.36) (2016,6.6) (2017,6.54) (2018,6.61) (2019,10.11) (2020,9.89) (2021,10.38) (2022,11.8) (2023,13.18) (2024,14.41)};
	\addplot+[ybar, fill=clr4, draw=clr4] plot coordinates {(2014,15.37) (2015,15.37) (2016,15.37) (2017,15.37) (2018,15.37) (2019,15.37) (2020,15.37) (2021,15.37) (2022,15.37) (2023,15.37) (2024,0.5)};
	\legend{\strut {Procurement, Grid Fee, Operation}, \strut {Renewable Energies Levy}, \strut {Other Levies}, \strut {Electricity Tax}}
	\end{axis}
\end{tikzpicture}
\caption{Average electricity price for new industrial consumers in Germany. Annual consumption 160~000 to 20~million~kWh, medium-voltage supply. Data source~\citep{BDEW}.}
\label{fig:electriticy}
\end{figure}
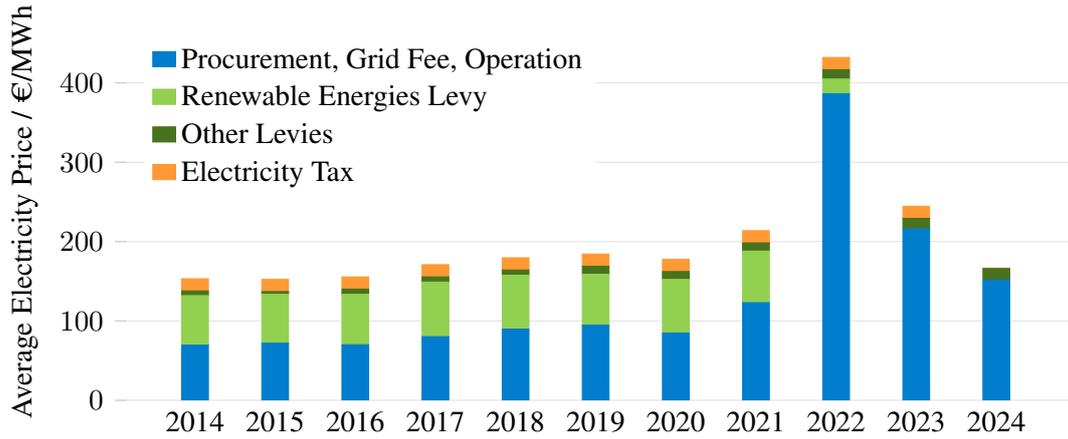

The fact that increases in computational performance necessarily come with a higher power envelope has several implications. Although power capping or power limiting can be used successfully (cf.\ Section~\ref{sec:powmgt}, and~\cite{Zhao_2023} and references therein), there is a delicate balance between power consumption and throughput. %\textcolor{red}{This paragraph and Figure~\ref{fig:electriticy} are new. Please check.} 
Electricity prices in Germany vary widely and are much higher for households ($\approx$~\euro{0.40}~per~kWh) than for industry ($\approx$~\euro{0.20}~per~kWh). Given the large amount of electricity consumed by HPC infrastructures, the prices centres pay fall into the latter category. Figure~\ref{fig:electriticy} shows how electricity costs for industrial consumers have evolved over the last few years, based on statistics published by the German Association of Energy and Water Industries~\cite{BDEW}. The costs are broken down into the categories of \textit{procurement, network charges, operation} (in blue), governmental levies to finance the development of the electricity network and new energy sources (in green)~\footnote{We show separately (in light green) the highest levy, the \textit{levy for renewable energies}, which was in force until 2021.} and the electricity tax (in orange). Until 2022, Germany's energy mix relied heavily on gas supplies from Russia, which stopped after the invasion of Ukraine and the subsequent international sanctions on Russian goods. This led to a sharp increase in electricity procurement costs in 2022, which the German government partially offset by reducing taxes and levies. Prices fell again in 2023, but even leaving aside the 2022 peak, the general trend shows an annual increase in electricity costs of $\approx$~\qty{3}{\percent}. These cost trends are based on averages, and it is important to note that the electricity prices paid by HPC sites vary considerably (from \euro{0.15}~kWh to \euro{0.29}~kWh for the sites in this study), as the institutions hosting the HPC centres have very different ways of purchasing electricity. 

The contribution of operational costs (including electricity and cooling costs) to the Total Cost of Ownership (TCO) of HPC systems hosted by our institutions ranges from \qty{12}{\percent} to \qty{50}{\percent}, where \qty{50}{\percent} means that running the system over 5 years costs the same amount of money as the initial purchase of the hardware (which typically includes a 5-year maintenance contract). Since the hosting site has little influence on the initial acquisition cost (only its negotiating skills), improving operational energy efficiency is the only strategy an HPC site can use to significantly reduce the TCO.

Legislation is a further driver enforcing energy efficiency measures in Germany. The Energy Efficiency Act \cite{enefG_dc}, which has been derived from the European Energy Efficiency Directive \cite{EUenergyefficiency}, mandates that (i) data centres that are currently in operation must reach a power usage efficiency (PUE) of less than or equal to 1.3 on a permanent basis by July 1st 2030 (regulations for newly deployed data centres are even stricter, cf.\ also Section~\ref{sec:dc}), (ii) data centers that go into operation starting July 1, 2026 must be constructed and operated in such a way that at least \qty{10}{\percent} of the waste heat is reused (this share grows to \qty{20}{\percent} for data centers going into operation on July 1, 2028).
The law also regulates air cooling temperatures, mandates the establishment of an energy management system, and contains further reporting duties.

\begin{table}[h!]
    \caption{Total power supply (in average) to each of the centers, broken then down on the relative consumption by three main components: infrastructure, compute, and storage. Notice that in the case of MPCDF the power consumption of storage cannot be measured separately and it is comprised in the compute part. }
    %\tinote{For TUD: only HPC (compute and storage) plus the infrastructure overhead for those systems, network is included in compute/storage, storage includes some other services (login, admin, viz)}
    %\tinote{For HLRS: Added separate numbers for the two different energy supply domains we have. HLRS I is the original setup of the data center from 2005. It supplies the cold water loop for air cooling by district cooling and dry coolers. HLRS II is the extension of the energy supply from 2011. It supplies water cooling with wet cooling towers and district cooling.}
    %\mfnote{KIT numbers only for Tier-2 system HoreKa}
    %\tfnote{Percentage for infrastructure is quite low since we have free cooling for JURECA DC and JUWELS Booster, and district cooling for JUWELS Cluster}
    \begin{threeparttable}
        \centering%
        \small%
        \begin{tabular}{p{1.5cm}<{\centering}p{6cm}<{\centering}p{3cm}<{\centering}p{2.5cm}<{\centering}p{2.5cm}<{\centering}}
        \hline
        \toprule
        \textbf{Centre}& \textbf{Average total power supply / \unit{\mega\watt}} & \textbf{Infrastructure} & \textbf{Compute}  & \textbf{Storage} \\
        \midrule
        DKRZ & 2.1 & \qty{13}{\percent} & \qty{80}{\percent} & \qty{7}{\percent} \\
        FAU  & 1.2 & \qty{15}{\percent} & \qty{80}{\percent} & \qty{5}{\percent}  \\
        %\hline 
        %HLRS I  & 0.46~avg & 21~\% & 54~\% & -~\%  \\
        %HLRS II & 3.1~avg & 12~\% & 86~\% & 2~\%  \\
        %\hline
        %HLRS & 3.56~avg & 13~\% & 82~\% & -~\%   \\
        HLRS & 3.6 & \qty{14}{\percent} & \qty{83}{\percent} & \qty{3}{\percent}   \\
        %JSC & {3.4~avg} & $<$9~\% & {82}~\% & 8.2~\%  \\ 
        JSC  & {3.4} & \qty{10}{\percent} & \qty{82}{\percent} & \qty{8}{\percent}  \\ 
        KIT  & 1.1 & \qty{10}{\percent} & \qty{82}{\percent} & \qty{8}{\percent}\\ 
        LRZ  & 2.8 & \qty{7}{\percent} & \qty{86}{\percent} & \qty{7}{\percent}  \\
        MPCDF & 4.7 & \qty{15}{\percent} & \qty{85}{\percent} & -- \\
        TUD  & 1.9 & \qty{15}{\percent} & \qty{78}{\percent} & \qty{7}{\percent} \\
        %TUD & 1.945 MW avg & 15~\% & 78~\% & 7~\% \\ 
        \bottomrule
        \end{tabular}
\end{threeparttable}
\label{tab:enbreakdown}
\end{table}

The HPC centres involved in this study currently operate (as of September 2024) production supercomputers with an average power envelope per site ranging from $\sim$\qty{1.1}{\mega\watt} (KIT) to \qty{4.7}{\mega\watt} (MPCDF). However, all sites have plans to upgrade their infrastructures for their next generation systems, e.g. the JUPITER exascale system will raise the bar to \qty{17}{\mega\watt} when added to JSC's existing computing infrastructure. 
It is interesting to note that, despite the relatively wide range of system sizes, the breakdown of power consumption across the various components within the data centre is quite homogeneous. We have collected in Table~\ref{tab:enbreakdown} average values, aggregating for each site the contributions of all systems it currently hosts. Most of the energy (\qtyrange{78}{86}{\percent}) is consumed by the computer itself, where unfortunately it is not possible to separate the contribution from the processing units, memory and network in the measurements\footnote{Some research projects provide these details, but reliable statistics for the large production systems are not yet available}. The second consumer (\qtyrange{7}{15}{\percent}) is the data centre infrastructure, which needs electricity to run uninterruptible power supplies (USV), pumps, climate machines (for air-cooled systems\footnote{A clear trend over the last 10~years shows that all sites are moving away from pure air-cooled systems towards direct liquid cooling (see section~\ref{sec:cool} for details), but some specialized systems still require air cooling.}), chillers (for cold-water cooled systems), dry coolers, etc. The remaining energy (\qtyrange{3}{8}{\percent}) is consumed by storage systems. In the following, we look in detail at the approaches actively applied or planned to save energy at the data centre level (section~\ref{sec:infra}) and the computer and storage systems (section~\ref{sec:syshw}).

\section{Infrastructure} % - [1500 words]}
\label{sec:infra} 
Data centres, whether for HPC or cloud systems, consume significant resources in the form of electricity and water, with a significant fraction being used for their infrastructure. 
The growth in computing power comes with higher computing density and heavier equipment, which has consequences for the hosting data centre. 
Cooling strategies are moving away from air-cooling to direct liquid cooling, utilising higher operating temperatures to enable free cooling. Adiabatic cooling expands the range of free cooling but on the other hand induces an increased demand for water. The following subsections describe these and further trends to reduce energy consumption at the data-center level.

\subsection{Electricity supply} %~230 words
\label{sec:electr}

To keep HPC operations sustainable, the growth in energy consumption needs to be compensated by increased usage of green energy so as to not increase the carbon footprint of HPC. Recent legislation in Germany requires data centres to continuously increase their energy efficiency and newly built data centers will have to source their energy completely from renewables by 2027, although quite a few data centres have been doing this voluntarily for quite some time already.

The fraction of renewables in the German grid already exceeds \qty{50}{\percent}~\cite{destatis_energy_24} and with the ongoing expansion of wind and solar power generation, this fraction is only set to increase.
While the intermittent generation of renewable electricity causes the grid stability to degrade, HPC centers could act as dynamic loads that increase their power consumption at times of high availability of green energy and reduce it when wind and solar generation is low.
This would not only stabilize the grid and help with the green transition, but also provide economic incentive for HPC operators as electricity prices can even turn negative when renewable production exceeds demand.

The technologies to implement such grid-demand-response schemes are available: power management capabilities as described in Section~\ref{sec:powmgt} can be used to modulate the power consumption of HPC systems dynamically and historical monitoring data of previous job executions could be leveraged to identify jobs with high power draw and schedule them at times of abundant energy.

\subsection{Data center} % ~420 words
\label{sec:dc}

The state of the art HPC data center design in Germany was in past years in almost all cases
done in a way where highly customized buildings with energy and cooling 
infrastructure tailored to 2-3~generations of HPC systems were planned and build from scratch. 
After about 10 years, infrastructure upgrades had to be put into place in order to keep up with the 
raise in energy consumption, cooling demands, and weight load of the HPC-systems to be installed. After several decades and upgrades,
delivering more energy and cooling into these customized 
buildings becomes more and more difficult as can currently be seen at the MPCDF, HLRS and JSC where new data centers are currently under construction or being planned for. While JSC has 
decided to set up a modular data center (MDC), MPCDF and HLRS are again setting up highly 
customized buildings. 
The MDC approach used at JSC, or generally spoken container-based datacenters as they are nowadays also used by commercial companies, allow for an exact match of the system requirements and data center provisioning without installation of over-capacities in infrastructure. In addition they significantly shorten planning and construction times, especially when it comes to standard IT. On the other hand, traditional building-based data centers, as planned at MPCDF and HLRS, provide a longer term solution that can be tailored to wider campus infrastructure plans to maximise the waste heat utilization.  
In all cases a new data center construction does not imply the old facility to be dismantled, but rather to be extended as in all cases the older buildings will remain in operation hosting smaller scale HPC-systems, cloud and storage systems, or regular IT-services. In those cases infrastructure demands did not increase as significant as for large-scale GPU accelerated HPC systems.

As of 2024, the trend to renew HPC data center infrastructure is not only forced due to system demands but also by the requirements of legislation~\cite{enefG_dc}. 
Legislative demands are fulfilled by both MDC and customized buildings, but they will become stricter in the future. After July 1st 2026, data centres (DC) shall be constructed and operated in such a way that their proportion of 
reused energy reaches at least \qty{10}{\percent}, while those starting operation one or two years later must
already reach \qtyrange{15}{20}{\percent} energy reuse
respectively. 
Taking these requirements into account, a foreseeable trend in Germany will 
probably be a change in data center operation away from a singular focus on the efficiency of 
the HPC-system or the DC towards integrated efficiency optimization of the HPC-system, DC and 
surrounding waste heat-absorbing district infrastructure.

\subsection{Cooling} % - [400 words]} -- reality 880
\label{sec:cool}

Compared to traditional air cooling, the adoption of direct liquid cooling (DLC) adds cost and complexity at the interface between IT systems and data centers.
However, due to the size and homogeneity of their compute clusters, HPC centers are ideally positioned to employ DLC.
Operating DLC loops at higher temperatures (\textit{warm} water, typically \qtyrange{30}{40}{\celsius}  inlet temperature) allows in Germany for year-round free cooling, eliminating the need for chillers and thereby reducing both capital and operating expenditure.
In addition, the ever-growing thermal design power (TDP) of today’s CPUs and GPUs mandates DLC for HPC centers aiming for highest performance.

The experiences with operating DLC systems have been largely positive.
In particular, fully integrated cooling solutions (w/o fans) have demonstrated excellent performance, achieving high delta T values ($>$\qty{10}{\kelvin}) and transferring more than \qty{95}{\percent} of the heat generated by HPC systems into the water. 

Some vendors upgrade air-cooled compute nodes with purpose-designed coldplates for liquid cooling.
This approach greatly increases the range of configurations to choose from, but, depending on the temperature, typically only captures about \qty{70}{\percent} of the heat in water as some parts of the mainboards as well as power supply units and networking equipment remain air-cooled.

Table~\ref{tab:DLC} summarizes information on water cooling adoption in German HPC centers, some of which have more than 10 years of experience operating DLC clusters.
Some aspects are common for most centers: All large storage systems and special compute node configurations that require more flexibility (fat nodes, test systems and similar) remain air-cooled but being supported e.g. by using water-cooled doors.
For air-cooling installations, there is little room for further innovation that could lead to improve e.g. efficiency and operational aspects~\cite{hackenberg_2016}.
Other thermal management technologies, such as immersion or two-phase cooling \citep{curtis2023}, are currently not utilized in HPC operations at our German sites.

\begin{table}[h!]
\caption{Direct Liquid Cooling (DLC) for Various German HPC Centers.}
\begin{threeparttable}
    \centering%
    \small%
    \begin{tabular}{c<{\centering}p{1cm}<{\centering}p{1.5cm}<{\centering}p{2cm}<{\centering}p{1cm}<{\centering}p{1cm}<{\centering}p{0.25cm}<{\centering}p{0.25cm}<{\centering}p{0.25cm}<{\centering}p{0.25cm}<{\centering}p{4.2cm}<{\centering}}
    \toprule
    \textbf{Site} & \textbf{DLC Since} & \textbf{\# Clusters} & \textbf{Power per cluster / \unit{\kilo\watt}} & \textbf{Inlet / $^\circ$C} & \textbf{Outlet / $^\circ$C} & \textbf{B} & \textbf{C} & \textbf{Q} & \textbf{H} & \textbf{Remarks} \\ \midrule
    DKRZ & 2015 & 1 & 2500 & 40 & 48 & \mycmark & \myxmark & \mycmark & \mycmark & building loop and cooling tower loop are combined (all water/glycol)\\ \midrule
    FAU & 2021 & 3 & 750, 250, 1000 & 30 & $>$40 & \myxmark & \myxmark & \myxmark & \mycmark & DLC only for CPU-only cluster (\#1) and hybrid CPU/GPU cluster (\#3)\\ \midrule 
    HLRS & 2019 & 1 & 3100 & 25 & 35 & \myxmark & \myxmark & \myxmark & \mycmark & \\ \midrule %planned: min. 32/42$^\circ$C, target 40/50$^\circ$C
    JSC & 2012 & 3 & 1580, 2046, 894 & 36 & $>$40 & \myxmark & \myxmark & \mycmark & \mycmark &  \\ \midrule
    KIT & 2016 & 1 & 1084 & 42 & 47 & \myxmark & \myxmark & \myxmark & \mycmark &  Only Tier-2 system HoreKa \\ \midrule
    LRZ & 2012 & 5 & 3000, 200-600 & 46 & \qty{52}{\celsius} & \myxmark & \myxmark & \myxmark & \mycmark & SuperMUC-NG: \qty{46}{\celsius} inlet, newer systems: \qty{43}{\celsius} \\ \midrule
    MPCDF & 2008\tnote{*} & 4 & 300-1000 & 16/40 & $>$20/47 & \myxmark & \myxmark & \myxmark & \mycmark & \qty{3}{\mega\watt} of spring water cooling (\qty{14}{\celsius}/\qty{20}{\celsius} max) \\  \midrule
    TUD & 2013 & 4-5 & $>$500 & 35 & 48 & \myxmark & \myxmark & \mycmark & \mycmark & \\ \bottomrule
    \end{tabular}
    \begin{tablenotes}
        \item \hspace{-0.7em}\textbf{B}: Biocide in direct cooling (building) water loop.
        \item \hspace{-0.7em}\textbf{C}: Corrosion inhibitor in direct cooling (building) water loop.%\nenote{is this column required since all centers are the same?}
        \item \hspace{-0.7em}\textbf{Q}: Experienced water quality challenges in direct cooling (building) water loop.
        \item \hspace{-0.7em}\textbf{H}: Heat exchanger/CDU between direct cooling (building) water loop and rack/system.%\nenote{is this column required since all centers are the same?}
        \item[*] Initial experiences since 1991 with various Cray systems.
    \end{tablenotes}
\end{threeparttable}
\label{tab:DLC}
\end{table}

For some centers
the adaptation to warmer cooling temperatures has been delayed by site-specific infrastructure legacy constraints such as local cooling networks.
However, most (all?) of those centers are in the process of transitioning to year-round (warm water DLC) free cooling for their compute clusters.
Typical operating regimes comprise multiple DLC clusters at \qty{2}{\mega\watt} total power, or more.

The design of the cooling loops needs to be carefully considered for DLC data centers, e.g., with respect to the number and placement of heat exchangers, while considering vendor interfaces.
Most centers in this study operate three loops, separated by heat exchangers: water/glycol for the (closed loop) cooling towers, water for the main building loop, and typically some vendor specific coolant for the rack/system level internal loop.
Eliminating heat exchangers can improve efficiency.
When building-loop water is circulated directly through compute nodes, maintaining high water quality (e.g., filtration and particle size) becomes critical, balancing efficiency with operational challenges (see below). 
Similarly, circulating building-loop water through cooling towers can improve efficiency, but preventing damage to the cooling loops is challenging when (unscheduled) system downtimes occur during freezing outside temperatures; alternatively, one large glycol loop can be operated at the expense of increased pump energy while saving heat exchanger installation and losses (cmp. DKRZ).
Alternative cooling strategies, such as using spring or lake water (cmp. MPCDF), may offer highly efficient cooling, but only for certain locations.

Concerning operational challenges for DLC installations, maintaining stable water quality is key.
The warm water of DLC loops stimulates bacteria / micro-algae proliferation.
Vendors address this issue with proprietary formula in the rack/system cooling loops.
For the building water loop, it may be managed through chemical treatment (biocide), or through maintaining extremely clean (e.g., deionized) water, with most sites in this study choosing the latter.
However, any water additive (glycol contamination, corrosion inhibitor or even biocide  
may eventually become food for some form of biology, creating an environment in which the additive balance must be continuously tuned.
Therefore, regular (external) water quality analysis or continuous monitoring (pH, conductivity, total organic carbon) is required in any case.

The use of Polyethylene/Polypropylene piping can offer financial and flexibility advantages, but may also lead to uncontrollable oxygen diffusion into the water and hence increase the risk of corrosion.
Particular care must be taken with metal mixes in the cooling loops and the associated chemical interactions that may strongly accelerate corrosion.

Another potential issue are water leaks, which do not happen frequently but have been seen in isolated cases, in particular located at the quick-connects inside the rack or at the connection between rack and secondary water loop.
Leakage sensors with automatic monitoring and alarm systems can mitigate the risk.

Operational challenges may also include potentially large load swings of the size of multiple MW at second-scales for big centers.
However, this issue is much more prevalent on the electrical side (sub-second time scales) and can be managed particularly well for chiller-free (direct water) free cooling loops with a reasonably sized buffer tank (water storage), enabling simple control loops and very steady operation.

Looking ahead, cooling temperatures may decrease, as recent guidelines from ASHRAE \citep{ashrae2021b} suggest.
This trend is understandable, because the density of compute power of currently more than 150~kW per rack and in the future 200~kW per rack and more can no longer be sustained through improved processor manufacturing processes or higher acceptable device temperatures, lower cooling temperatures allow a larger heat dissipation.
However, it may necessitate the reintroduction of chillers, which would be a significant setback in terms of both efficiency and environmental impact.

\subsection{Heat reuse} % - [400 words]} -- reality 477
\label{sec:reuse}

Waste heat reuse is among the most effective means to improve sustainability of data center operations.
Due to their increased cooling temperatures, DLC installations are particularly well-suited for efficient heat reuse.

Any such project strongly depends on the local environment of the data center, because the waste heat needs to be consumed locally.
Our survey in Table~\ref{tab:HeatReuse} shows energy reuse factors ranging from 0 to 20\% for these DLC-enabled German HPC data centers.
The current potential is not fully utilized, even tough highly ambitious plans exist at several sites. 

In particular, year-round high levels of heat reuse are difficult to achieve, because heat generation of HPC centers often exceeds the heat demand of surrounding buildings even during winter.
Usually, the most promising large scale heat sink are city-wide heat networks.
In Germany, these usually operate at temperatures close to \qty{100}{\celsius}, requiring heat pumps to inject low-temperature data center waste heat.
Operating heat networks at temperatures that would allow for HPC waste heat to be reused w/o heat pumps would be technically viable and overall more energy efficient.
However, retrofitting all consumers (i.e., residential heating) is impractical within the typical life span of HPC data centers.

Political regulations have been put in place to mandate 20\% heat reuse for data centers that go into operation in 2028 an later, but this would strongly influence the range of viable building sites and potentially harm data center business in Germany; the practical implications of this legislation are therefore still fiercly debated.

\begin{table}[h!]
\setlength{\abovetopsep}{2pt} % AH: for some unknown reason to me, there is too little space between the table caption and the toprule
\caption{Heat Reuse for Various German HPC Centers}
\begin{threeparttable}
    \centering%
    \small%
\begin{tabular}{c<{\centering}p{0.8cm}<{\centering}p{1.4cm}<{\centering}p{0.8cm}<{\centering}p{0.5cm}<{\centering}p{1.8cm}<{\centering}p{1cm}<{\centering}p{1.3cm}<{\centering}p{0.8cm}<{\centering}p{0.5cm}<{\centering}p{2.5cm}<{\centering}} %p{4cm}<{\centering}|}
\toprule
\textbf{Site} & \multicolumn{5}{c}{\textbf{Currently in Operation}} & \multicolumn{5}{c}{\textbf{Planned for Future}} \\ %& \multirow{2}{*}{\textbf{Remarks}}  \\ \hline
\cmidrule(lr){2-6}
\cmidrule(lr){7-11}
 & \textbf{Since} & \textbf{avg./max.} & \textbf{ERF} & \textbf{HP} & \textbf{Consumer} & \textbf{Year} & \textbf{avg.} & \textbf{ERF} & \textbf{HP} & \textbf{Consumer} \\ 
\midrule
DKRZ & 2019 & \qty{340}{\kilo\watt} / \qty{760}{\kilo\watt} & 0.20 & \myxmark & neighboring labs & \multicolumn{5}{p{8cm}}{Possibly extent current installation with next HPC system in 2027} \\ \midrule % 3 GWh/year
FAU & - & - & - & - & - & \multicolumn{5}{p{8cm}}{Considerable reuse starting 2029} \\ \midrule % 'Considerable' reuse starting
HLRS & 2011 & \qty{2}{\kilo\watt} / \qty{4.2}{\kilo\watt} & 0.009 & \mycmark & office heating & 2028 & \qty{5.5}{\mega\watt} & 0.5 & \mycmark & campus district heating \\ \midrule
JSC & 2024 & 0 / \qty{1800}{\kilo\watt} & - & \myxmark & neighboring buildings & $>$2025 & \qty{3.6}{\mega\watt} & 0.24 & \mycmark & campus building \\ \midrule %  heat pumps and capacity extension $>$3 MW reuse, heat pumps included 
KIT & 2016 & \qty{30}{\kilo\watt} / \qty{200}{\kilo\watt} & 0.04 & \myxmark & office heating & \multicolumn{5}{p{8cm}}{future plans (2030+) for $>$\qty{2}{\mega\watt} heat reuse, potentially campus district heating} \\ \midrule
LRZ & 2014 & \qty{240}{\kilo\watt} / \qty{600}{\kilo\watt} & 0.06 & \myxmark & office heating & \multicolumn{5}{p{8cm}}{Potentially extend to neighbouring campus }\\ \midrule %2.1 GWh/y office use %newly built offices on campus %Total consumption (full data center) 33.6 GWh, ERF=0.06 %400 kW for office heating
MPCDF  & -  & -  & - & - & - &  \multicolumn{5}{p{8cm}}{future plans (2030+) for up to \qty{70}{\percent} heat reuse on campus (offices and labs) and potentially city district network}\\ \bottomrule 
TUD & 2016 & \qty{205}{\kilo\watt} / \qty{360}{\kilo\watt} & 0.12 & \myxmark & local heating network & 2025 & \qty{2}{\mega\watt} & 0.77 & \mycmark & city district network  \\ \midrule % 
\end{tabular}
    \begin{tablenotes}
        \item \hspace{-0.7em}\textbf{avg.}: Average heat load over 12 months of a reference year.
        \item \hspace{-0.7em}\textbf{ERF}: Energy Reuse Factor (reused heat / total heat)
        \item \hspace{-0.7em}\textbf{HP}: Heat Pump
        \end{tablenotes}
\end{threeparttable}
\label{tab:HeatReuse}
\end{table}

So far, no noteworthy projects that feed HPC data center heat into industry processes with year-round demand have emerged.
Table \ref{tab:HeatReuse} shows that, for now, the most promising approach is using large heat pumps to increase the temperature to the levels required by district heating networks of cities or very large campuses (usually around \qtyrange{70}{90}{\celsius}).
At TUD, such a system will be fully operational in Q1 2025.

Ideally, heat reuse becomes the primary heat rejection mechanism, and cooling towers of the DLC loop will be relegated to a backup role. 
This may alter the cooling loop design in the future: the focus shifts from \textit{cooling-first} to \textit{heat-reuse first}, with a greater emphasis on strategically placing heat exchangers to optimize for efficient energy recovery, while the efficiency of the \textit{backup} cooling becomes largely irrelevant.

Other ambitious projects exist, e.g., for the new HLRS data center, water-cooled CO$_2$ compression chillers are planned to make lower temperature cooling circuits, which are also necessary for air cooling and climate control, available for waste heat utilization.
An innovative concept to use waste heat to drive adsorption chillers for cold water production has been evaluated by LRZ with convincing results~\cite{wilde2017}, but recent trends to lower water temperatures in DLC rendered this technology inefficient for data center use.

\section{System hardware} % - [2000 words]}
\label{sec:syshw}

The physical limits to ever-shrinking structure sizes and the associated manufacturing challenges are slowing Moore's Law~\citep{huang2015}. The computing industry, and the HPC community in particular, are now trying to increase performance through diversification and specialisation~\citep{milojicic2021, Shalf_2020}.

\subsection{Processor Heterogeneity} % (MSA, chiplet integration, etc). - [1000 words]}
\label{sec:hwhet}

Processing units are the most power hungry part of an HPC system. Since the introduction of \textit{cluster computing}, HPC systems have predominantly been built using high-speed networks to interconnect general purpose processors (CPUs) from the server market~\citep{beowulf1995}. Since around 2010, such clusters have started to integrate GPUs and other accelerators to increase performance per Watt~\citep{carabano2013}. Compared to CPUs, accelerators pack a very larger number of simpler cores or execution units per area, sacrificing single-thread performance for higher overall throughput for parallel computing. They further increase energy efficiency by specialising their architecture to solve specific computations, relying on CPUs mainly for service tasks. The result are higher FLOP/Watt ratios for accelerated systems, at least when considering arithmetic intense applications such as the High Performance Linpack benchmark~\citep{top500}.

The goal of heterogeneous HPC systems is to offer users different compute devices, allowing them to select those best suited for their applications. Depending on how tightly the CPU and accelerators are tied to each other, different heterogeneous system architectures are possible. The two extremes can be classified as \textit{monolithic} and \textit{modular} system architectures. 
A monolithic system packs many different kinds of processing units inside each node creating a system-wide homogeneous cluster (all nodes look the same) made of highly heterogeneous nodes (various compute devices inside the node). This approach minimises the communication latency between host and accelerators~\citep{schulz2021}, but it can lead to low resource utilization and higher energy consumption. This can be mitigated by applying advanced scheduling techniques --~to efficiently co-schedule complementary jobs and keep all accelerators busy~--, and dynamic power management --~to switch off all unused devices~\citep{suarez2022}. 
Modular systems, on the other hand, are system-wide heterogeneous supercomputers made of a diversity of homogeneous clusters (modules) interconnected to each other via a high speed network~\citep{suarez2019, suarez2022}. Each module has a specific node architecture, with ideally only one kind of processing unit. This resource disaggregation makes it easier for the scheduler to reserve only the kind and amount of nodes needed by each application without blocking resources for others, but relies on a coarse-granular partition of the application codes to ensure that performance does not suffer from the larger inter-module communication latency~\citep{kreuzer2021}. It is worth noting that resource disaggregation can be beneficial for compute, but also for memory devices~\citep{michelogiannakis2022, aguilera2023}. Interconnect technology is already supporting the creation of memory pools accessible via the network with network protocols such as CXL~\citep{gouk2023}.

In the German landscape, all HPC centres included in this study host heterogeneous computers combining CPUs and GPUs. These are either operated as independent systems (e.g. FAU, LRZ) or partitions or modules in integrated machines (e.g. DKRZ, JSC, MPCDF). The CPU systems/partitions consist of dual-socket nodes with x86 CPUs (Intel or AMD CPUs), sometimes organised in sub-partitions with different memory configurations (e.g. standard and large memory capacity). Only the planned JUPITER Booster at JSC will utilize non-x86 Arm CPUs in the NVIDIA Grace-Hopper configuration. The GPU systems currently all employ NVIDIA GPUs, mostly NVIDIA Ampere~(A100) or NVIDIA Volta~(V100), depending on the year of deployment, while a number of NVIDIA Ampere A40 are also installed, e.g. at FAU. Most recent and planned installations are foreseen with AMD MI300 GPUs (MPCDF and HLRS). All centres use Slurm as their batch scheduling system~\citep{slurm}, supporting heterogeneous jobs and allowing applications to allocate resources across different compute partitions.

More hardware heterogeneity exists in small test platforms and also in cloud infrastructures operated by the German centres, e.g. MPCDF, TUD, KIT, and JSC all host a small number of different devices, e.g. AMD GPUs, Graphcore, FPGAs. In fact, with Moore's Law reaching its limits, a variety of accelerators addressing different use cases are emerging, many of which are targeting the vast AI market~\citep{matsuoka2018}. GPUs are still the most widely used accelerators in HPC, but new ASIC designs are attracting attention (e.g. TPUs~\cite{tpu}, Graphcore~\cite{graphcore}, Cerebras~\cite{cerebras}, Groq~\cite{graphcore}, SpiNNaker2~\cite{spinnaker}, etc.). A common trend across all these accelerators is a strong focus on lower precision arithmetic. This shift is justified by the fact that the number of operations per second a processor can perform is inversely proportional to the number of bits required to encode its floating point numbers. The same applies to energy efficiency, which is greater the lower the arithmetic precision used. However, lower precision arithmetic comes at the cost of higher uncertainty and less reproducible results. While this is acceptable for AI training applications, it remains to be seen how many traditional HPC codes can make the leap. 

Hardware heterogeneity goes down from the node into the package level. Chiplet-based processor designs combine different kinds of technologies and manufacturing processes to achieve higher performance without continuously growing the die area. For example, AMD EPYC Rome mounts CPU and I/O chiplets on top of an interposer to create an integrated processor with more than 100 cores~\citep{suggs2020, amdEpyc}. Chiplet-based designs offer more flexibility and customization, as they allow to define for a processor generation a diversity of SKUs with different numbers of CPU, accelerator, memory-, and I/O components, keeping the rest of the architecture more or less untouched. Interesting trends are also the increasing number of companies developing CPUs for the server, HPC and AI markets using the Arm Instruction Set Architecture (ISA), (e.g. Fujitsu~\cite{sato2020, fujitsu}, Ampere~\cite{ampere}, Apple~\cite{apple}, SiPEARL~\cite{sipearl}, NVIDIA~\cite{nvidia}), and the growing community around the open source RISC-V ISA~\cite{riscv}. The European projects \emph{EPI}~\cite{epi}, \emph{EUPILOT}~\cite{eupilot} and \emph{EUPEX}~\cite{eupex}, for example, work on Arm-based CPUs and a variety of RISC-V-based accelerators, targeting excellent energy-efficiency. These and upcoming R\&D projects in Europe target future chiplet-based integration combining some of the developed processing technologies~\cite{eurohpc-dare}.

However, all the above discussed hardware heterogeneity comes at the cost of higher programming complexity, since most accelerators require specific programming models and force application codes to be ported and refactored (see Section~\ref{sec:prog}). 

\subsection{Storage} % - 400 words } - real 473 words
\label{sec:stor}

Storage systems consume a smaller portion of total energy compared to the processing part. The German HPC centers included here offer 100s of Petabytes of storage space using a variety of file systems, in a mix of NVMe, HDD and Tape. Nonetheless, these HPC centers see no more than \qty{8}{\percent} of total power being consumed by storage, with some as low as \qty{3}{\percent} (see Table~\ref{tab:enbreakdown}). On the other hand, storage systems tend to be unique, are tightly coupled with the HPC systems, have higher availability requirements, and represent higher risk on outages. Therefore, any power saving techniques need to be well tested and adhere to further stricter limitations.

There are three aspects to consider when analyzing power consumption and evaluating energy-saving strategies for storage systems: i) energy consumption for data at rest or idle, ii) energy consumption for data access, and iii) effect of storage components on energy consumption in other parts of the system.

Storage systems are sized for both capacity and performance, and this in turn determines a baseline power consumption. This baseline, or energy consumed while the data is at rest, represents the largest share of a storage system's overall power usage, and it is hardly possible to reduce it. For example, the strategy of switching off idle components used in computing clusters is not feasible for a storage system without large data migrations or risking data loss. Overtime, storage systems have benefited from the advances in semiconductor technologies, leading to drives with larger capacity and a smaller set of servers required for the same number of drives. This has led to newer, bigger, faster storage systems consuming less energy. For example, the JSC's storage cluster JUST5~\cite{JUST} consumed an average of \qty{157}{\kilo\watt} power offering \qty{70}{\peta\byte} storage capacity and $\sim$\qty{400}{\giga\byte\per\second} bandwidth, while the newly installed JUST6 consumes about \qty{108}{\kilo\watt} ($0.7\times$) offering \qty{150}{\peta\byte} ($2.1\times$) and $\sim$\qty{600}{\giga\byte\per\second} ($1.5\times$). Future cheaper high capacity SSDs replacing HDDs shall reduce power consumption per storage unit~\cite{Tomes2017} but require a larger number of servers to leverage.

Regarding the second aspect (energy consumption for data accesses), storage systems under load consume only slightly more energy than at idle. For example, observations on JSC's, HLRS's as well as LRZ's storage clusters show that power consumption during operation does not vary by more than \qty{12}{\percent} from the highest peak\footnote{For JSC's storage cluster (IBM Storage Scale) the variation observed is \qty{12}{\percent}, for HLRS's (Lustre) \qty{9.4}{\percent}, and for LRZ's (IBM Storage Scale) \qty{8.8}{\percent}.}.
This observation is supported by the fact that storage appliances typically run their corresponding processors in performance mode and disks are not turned off, whereas idling HPC compute nodes could be turned off or at least put into powersaving modes. At the same time, vendors only commit to a certain performance when the appliances power settings are not touched.

The third aspect is the effect of the storage system on the energy consumption of other components, for example, idling compute nodes due to slow I/O. There are strategies which could help mitigating this effect: accurately sizing the storage system, adding local storage, improving caching, asynchronous I/O, etc. However current observations show that the peak performance of storage systems is rarely reached during production \cite{maloney2024}, suggesting that optimizing applications' I/O behaviour would be a better approach. Alternatively, deploying NVMe based storage systems, which are less sensitive to I/O patterns, could automatically reduce idle times on I/O.

\subsection{Power management} % (dynamic voltage management, system-portion shut downs, etc) - 500 words}
\label{sec:powmgt}

As HPC systems are growing ever larger and becoming increasingly power hungry, managing their power consumption is important for their (energy-efficient) operation. All modern CPUs and GPUs provide software interfaces to either limit their power consumption directly or by controlling the operating frequency and voltage (dynamic voltage frequency scaling – DVFS) and hence the power consumption indirectly.

While higher clock frequencies in general translate into higher performance of these devices, they are usually detrimental to their energy efficiency: depending on the characteristics of a particular workload, the highest frequency may not always yield the shortest runtimes, e.g., because CPU cores are waiting for slow memory transfers. In particular, memory-bound workloads typically do not benefit from higher frequencies and can be executed at lower frequencies with negligible impact on performance and runtime.
However, as the energy consumption of an application run is the product of its runtime and average power consumption, a reduced frequency can yield energy savings for such workloads. In contrast, the energy efficiency of a compute-bound workload may benefit from higher frequencies as a potential increase in average power consumption may be compensated by shorter runtimes~\cite{auweter2014}.

Multiple approaches exist to leverage such energy saving schemes in production HPC environments: the simplest approach is to let the user select a frequency as they know the characteristics of their applications best. % and hence are able to pick a suitable frequency. 
Batch schedulers such as Slurm provide control parameters for job scripts that allow users to set the frequency for their jobs. However, a single frequency may only be optimal for some phases of an application run, but inadequate for others~\cite{Corbalan2020}. To deal with such situations, more sophisticated runtime systems such as EAR~\cite{Corbalan2020} (deployed at LRZ), GEOPM~\cite{Eastep2017} or HPE PowerSched \cite{powersched} (deployed at HLRS) are required. They dynamically set optimal frequencies at runtime and use online monitoring to assess the current characteristics of the workload and hence can accommodate different execution phases.

Another important field for power control are over-provisioned systems, i.e., systems that have a peak power consumption above the limits of their supporting electrical or cooling infrastructure. In day-to-day operations, HPC systems typically draw around 65\% of their peak power consumption. So, it can be cost-effective to design the supporting infrastructure for this lower power limit. However, this requires the system to stay below this limit at all times. The simplest approach here would be to enforce the same static \textit{power cap} on each individual compute node of the system to ensure that the total power limit is not exceeded. However, this may leave compute capacity stranded as the power characteristics of the various workloads running on the system could be very different. Here, again, a dynamic approach as implemented by HPE PowerSched and deployed at HLRS can assign dynamic power limits to individual jobs to best utilize the supporting infrastructure and maximize system throughput.

Even though HPC systems strive for a high utilization, idle power consumption is also of concern as certain situations (backfilling, maintenance) can lead to entire compute nodes being fully idle for longer periods of time.
Recent CPU-based systems exhibit a wide range of power consumption in idle~\citep{Troepgen_2024_SPEC} which puts a new emphasis on considering idle power consumption for procurement and its optimization during operation~\citep{2024_Ilsche_Idle}.

\section{Monitoring system} % - 1800 words}
\label{sec:monit}

The energy efficiency of a system as complex as an HPC infrastructure can only be optimised if it is first properly quantified, which requires careful monitoring of all the components that contribute to energy consumption.
The size and complexity of HPC center environments place high demands on the collection and storage of metrics.
Table~\ref{tab:monitor} showcases the range of requirements of the represented German HPC monitoring systems with up to 8 million total metrics and up to 10 updates per second for each time series.
At the same time, agents collecting data on the compute nodes should not interfere with regular production codes.
Batch jobs with complex sets of used resources are typically not directly supported in generic monitoring stacks.
Because performance is a key focus of HPC systems, continuous cluster-wide measurement of hardware performance counters provides critical data to judge the efficiency of batch jobs.
In addition, HPC is a multi-user environment with different levels of trust, therefore strict security and access control are required, especially if unprivileged HPC users have access to the monitoring system.

\begin{table}[h!]
\sisetup{range-phrase=-}
\caption{Characteristics of monitoring installations on German HPC data centers.}
\begin{threeparttable}
    \centering%
    \small%
    \begin{tabular}{p{1.1cm}<{\centering}p{3.4cm}<{\centering}c<{\centering}p{3.7cm}<{\centering}p{5.1cm}<{\centering}}
    \toprule
    \textbf{Centre} & \textbf{\#Metrics} & \textbf{Intervals [\unit{\second}]} & \textbf{Long-term Storage} & \textbf{Tech Stack} \\
    \midrule
     DKRZ & 960k (42/node + 9/socket + 2/core) & \numrange{1}{60} & Meta: indefinite, Metrics: 6 months & Collectd, Prometheus, ClusterCockpit, Elasticsearch, Grafana \\ \midrule
     FAU & 660k (8/node + 7/core + 3/socket + 6/GPU) & \num{60} & Job data including metric dataL indefinite & ClusterCockpit, NATS, Grafana, Munin \\ \midrule
     % Removed from FAU for space: (cc-metric-collector, cc-metric-store, cc-backend)
     HLRS & 500k & \numrange{1}{120} & 8 weeks - indefinite & collectd, Telegraf, LDMS, Kafka, BarrelEye, TimescaleDB, Influx, Opensearch, MS SQL, Grafana \\ \midrule
     JSC & 3.4M ($\sim$600/node) & \num{60} & 14 weeks - indefinite & Prometheus, Promtail, Loki, Grafana, LLview \\  \midrule
     % Removed from JSC for space and easier overview
     % Split "nodes: 3M ($\sim$600/node) datacenter: 400k"
     % 2) Defailts for Prometheus (node, dcgm, ipmi)\_exporter
     KIT & 115k (144/node) & \num{30} & {3 months} & {JobMon, ClusterCockpit, InfluxDB, } \\ \midrule
     % KIT, TI: replaced cc-metric-collector with ClusterCockpit for simplicity
     MPCDF & 100k (40/node) & \numrange{3}{240} & indefinite & hpcmd, rsyslog, Splunk, PDF~reports\\ \midrule
     LRZ & 8M (56/node + 12/core) & \numrange{0.1}{30}& 30 days, sub-samples indefinite & DCDB, MQTT, Cassandra, Grafana\\ \midrule
     TUD & 330k (24/node + 4/core + 4/GPU) & \numrange{0.25}{30} & indefinite & MetricQ, RabbitMQ, PIKA, Grafana \\ \bottomrule
     \end{tabular}
\end{threeparttable}
\label{tab:monitor}
\end{table}

\begin{figure}[h]
\begin{center}
\includegraphics[width=18cm]{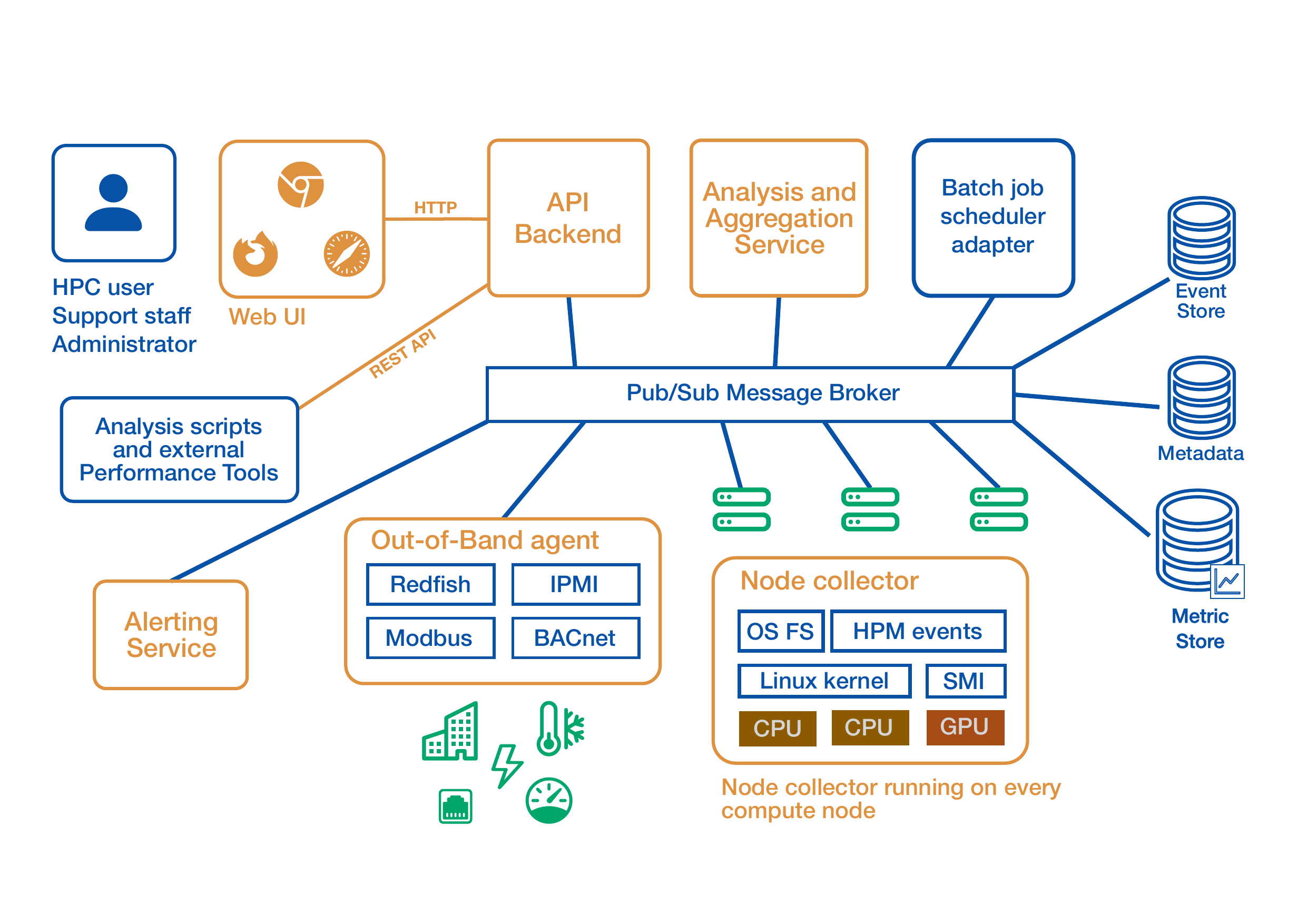}
\end{center}
\caption{Typical components of a monitoring setup in HPC-Cluster environments}\label{fig:monitoring-architecture}
\end{figure}

Figure~\ref{fig:monitoring-architecture} shows typical components of a monitoring stack in HPC environments.
For data collection, a node collector running on all compute nodes retrieves metrics from various interfaces.
These values are then either sent to or fetched by a communication component, typically a message broker, for processing, storage, and visualization.
Metrics from infrastructure components such as power distribution, cooling infrastructure, file systems, and network are retrieved out-of-band using different communication protocols.
Because the central notion of work on HPC systems is a batch job, the monitoring system must integrate with the batch job scheduler (e.g. PBS, LSF, Slurm, etc.).
Typically, a web interface provides access to the monitoring data.
Different user groups (HPC users, project managers, support staff, administrators) may have different data access restrictions and require specific views.
Automatic analysis of jobs and general system attributes can be used to focus attention on where immediate action is required, including automatic alerting (e.g. via e-mail).

\subsection{Data collection and organization} % in-band/out-of-band }%(which sensors, which tools) - [500 words]}
\label{sec:datacol}

Monitoring of HPC systems and their supporting infrastructure require the collection of relevant metrics from a wide variety of different sources: the compute nodes, the network infrastructure, the storage system, and ideally also the building infrastructure.
There is no common API to collect such data and multiple different interfaces need to be queried for different sub-components.
For example, within a compute node, there are many sub-components such as CPUs, GPUs, and network interfaces, each of which has their own and often vendor-specific interfaces.
Some sub-components even have multiple interfaces, depending on the type of metric to be collected.

Most metrics for compute nodes need to be collected in-band, i.e., by a daemon running on the compute node itself (the node collector).
This daemon will perturb other processes and great care must be taken to keep the overhead minimal in order to not interfere with the user applications running on the node.
In particular the measurement of hardware performance monitoring (HPM) metrics involves a significant overhead~\cite{Gruber2014}.
A common practice is to benchmark a very demanding application, as e.g. the Linpack benchmark, with and without monitoring, to proof that there is no measurable influence.
Even if there was a small overhead, the benefit of a monitoring system that allows to detect pathological jobs and gives support personnel and users critical insight into job runs, usually overcompensates any potential cost.
Also, for some metrics, such as HPM, exclusive access is required, which may conflict with other tools, e.g. for performance profiling.
There exist multiple generic node collectors that already implement the data collection, namely collectd~\cite{collectd}, cc-metric-collector (part of ClusterCockpit)~\cite{Eitzinger2019}, and Telegraf~\cite{telegraf}.
The access to HPM metrics can be provided via the PAPI~\cite{browne2000} and LIKWID~\cite{hager2010} libraries, or using the Linux perf\_event interface.

Other components such as network switches, power distribution units, or the data centre building infrastructure can be queried out-of-band from a management node via the network without involving the compute nodes.
Rather than minimal perturbation, such software is optimized for covering a large number of devices and high message throughput.
While there is a trend, at least for IT components, towards the RedFish standard to make monitoring data available via a well-defined and well-documented API out-of-band, other decades-old protocols and interfaces such as SNMP and IPMI or Modbus and BACnet are still widespread.
Many of these out-of-band interfaces have high latency and are hence not suitable for high frequency telemetry.
Also the point in time of the actual measurement may be not exact if it is not provided by the device itself but by the querying daemon.
Some of the existing node collectors also support accessing external components via out-of-band plugins.

Collection of monitoring data needs careful planning. The number of metrics available, particularly on compute nodes, can be overwhelming~\cite{netti2019}. For some metrics it may not always be clear what exactly is being measured and where. The collection interval needs to be carefully chosen, based on the update frequency of underlying sensors, frequency of change, and expected data volume.
Most importantly, a versatile naming scheme for metrics needs to be chosen that works for all metrics regardless of their domain.
For example, HPC systems are typically hierarchically organized (e.g. partitions, racks, nodes, sockets, CPUs, and cores), while their supporting infrastructure may not be.
Once the naming scheme has been settled and the metrics to be collected chosen, it is also important to record meta information for all metrics: what exactly are they measuring, unit of measurement, update frequency, etc.
The ClusterCockpit project proposes a generic data structure specification as JSON schema, that also includes metric lists~\cite{ClusterCockpitSpec}.

\subsection{Data aggregation, processing and storage} % (how to manage data volumes) - [500 words]}
\label{sec:dataagg}

Processing the collected data can be challenging due to the number of metrics, varying update rates, the number of monitored nodes, and the overall event/data rate.
Generic solutions for time series data that cover ingestion, processing, storage, retrieval, and often specific data collection agents can be used in the HPC data-center monitoring context.
InfluxDB~\cite{influxdb}, for example, ingests data from clients via a simple line protocol, particularly in conjunction with Telegraf.
Prometheus~\cite{prometheus}, another popular solution with its origins in cloud computing and clearly focusing on collecting metrics, uses a centralized server that pulls data from clients. A hierarchical setup can be used to address scalability demands in case of a large amount of clients or metrics. JSC for instance uses Prometheus to collect all metrics for its HPC systems, storage servers and their surrounding infrastructure. % TODO Explanation
A pull-based mechanism enables centralized control of update rates and simplifies client configuration.
In contrast, the push method offers easier network access configuration, particularly for clients in restricted networks, and incurs lower latency between data collection and processing.

Additionally, there are several solutions focusing on HPC data-center monitoring, namely ClusterCockpit~\cite{Eitzinger2019} (FAU), DCDB~\cite{netti2019} (LRZ), HPCMD~\cite{Stanisic2020_HPCMD} (MPCDF), and MetricQ~\cite{Ilsche2019_metricq} (TUD).
In these systems, the data collecting agents publish data to messaging systems (NATS, MQTT, rsyslog, RabbitMQ respectively), which enables flexible consumption for storage, transformation/aggregation or live analysis.
NATS and MQTT are more lightweight whereas RabbitMQ allows metric-identifier-based routing of messages, enabling full decoupling of clients from consumers.

Given the high cardinality and in some cases high individual data rate, scalable time series storage is particularly challenging.
The specific solutions differ in their implementation\footnote{All solutions provide interchangeable storage back-ends, listed are the defaults}:
DCDB uses Apache Cassandra, a scalable, open-source NoSQL database.
ClusterCockpit implements a custom in-memory database using fixed-duration ring-buffers for fixed-rate time-series data.
Long-term data in ClusterCockpit is aggregated by job and stored in a job archive.
MetricQ leverages the Hierarchical Timeline Aggregation (HTA) concept, which aggregates data in different levels on ingestion.
This scheme allows efficient and complete aggregate queries over large amounts of data points.

\subsection{System wide data analysis } %(formerly Visualization) - 400 words}
\label{sec:datavis}

Efficient system operation relies on data --- including energy monitoring data --- to provide actionable insights.
For day-to-day operations, the data is often presented in dashboards~\citep{Bates2016_dashboards}, but the information is also used in form of reports for strategic decisions, e.g., capacity planning.

A widely-used (for example at DKRZ, JSC, LRZ) web-interface for visual data presentation is Grafana~\cite{grafana}, which supports many built-in data sources (e.g., InfluxDB and Prometheus) but also custom data sources. It is highly customiszable, offering means to allow a "birds-eye view" of a full system down to single rack or even node analysis, where multiple data sources allow the integration of e.g. additional infrastructure information.
TUD uses the MetricQ WebView interface for explorative interactive monitoring data visualization.
This custom implementation supports low-latency zooming and displays time-series statistics using a min/max/average band for each metric.
ClusterCockpit and PIKA~\cite{Dietrich2020_pika} provide own custom web-interfaces, which allows for UI elements and authentication mechanisms that might be difficult to implement in generic solutions.
In ClusterCockpit a special emphasis is put on a flexible search, filter and sorting UI with high plot render speed.

Monitoring data is also used for alerting, i.e., directly notifying operators about anomalies, e.g., power consumption that is outside of the expected range.
For example, the TICK-stack around InfluxDB, includes Kapacitor~\cite{kapacitor}, which pushes alerts to handlers, including, email or instant messengers.
Prometheus provides an a dedicated Alertmanager service, which queries Prometheus for metrics and can render different alerts across multiple different channels.
As another example, TUD uses a custom agent subscribed to the stream of metric monitoring data that checks if values exceed thresholds or are not updated in expected intervals.
This agent then pushes alerts via the NSCA protocol to a Centreon installation that covers the general IT-Infrastructure of the university and handles notification via E-Mail.

\subsection{Job-specific analysis and insight extraction %(data correlations, application fingerprinting, ML-ansätze, etc.) - 400 words
}
\label{sec:datainsight}

Job-specific analysis refers to detecting job features or classifying jobs, which allows for more meaningful cluster-usage statistics and detecting faulty jobs or jobs with high optimization potential.
One use case is to detect the actual application or software stack used in a job.
This allows to gain insights about how a system is used by a specific application and in turn for a better tailoring of the system hardware to the application mix.

For classifying job performance or energy behavior often simple job statistics (called job footprint in the following), as average or maximum for a set of metrics, are used.
This job footprint is often based on metrics that describe the resource utilization of a job, as e.g. flop rates, memory bandwidth, load, memory capacity used, and file and network IO utilization.
There are attempts to formulate simple rules based on this job footprint and use threshold-based binning of jobs to classify them.
Due to the varying length and number of resources used, machine learning (ML) on the time series data is feasible but challenging.
There are ongoing investigations how ML can be used to provide additional insights, but none of the German sites in this study are using this in production at present.
A use-case for classification is automatic notification of support personnel or users, that a job or user requires attention.
Some centers %\textcolor{red}{which centres?} 
already use script based notification of users for faulty jobs, at many of the centers participating in this paper detection of such jobs or users is still performed by manual monitoring of the job data.

For administrators and support personnel, overview of and insight into the data after the classification can be obtained by routinely inspecting graphical dashboards that list recent problematic jobs.
JSC uses LLview to monitor the overall system utilization and provide detailed insights into specific jobs via detailed job reports available to end-users~\cite{llview}. It significantly simplifies spotting under-utilized resources without the need for additional instrumentation of the used codes. LLview is re-using the data provided by Slurm and by the above mentioned system monitoring Prometheus, where the later is continuously extended to support additional metrics of particular interest, e.g. GPU utilization.
MPCDF uses Splunk dashboards that identify jobs with harmful behaviour (e.g. issuing file open and close operations at high frequency on the shared file system) or wrong resource requests (e.g. not using GPUs on GPU nodes).
Based on similar logic, HPC users get notified for individual jobs via the \texttt{stdout} stream at the end of a job about obvious inefficiencies.
Another use-case of job footprints and classifications is cross-job analysis to investigate similarities or differences across user groups, application classes, and applications~\cite{maloney2024}.
Statistics about the resource utilization of jobs is very important for preparing cluster procurements, decide on what systems are suited for a user or application class, and to optimise scheduling and resource management policies.

\section{Scheduling and resource management} % (dynamic scheduling, node sharing, etc.) - [1000 words]}
\label{sec:sched}

Energy efficiency is as much about keeping the power consumption of the system and its infrastructure as low as possible, as it is about ensuring that hardware resources are optimally utilized to deliver the maximum overall throughput of application results. Resources are shared between many users simultaneously executing their codes on HPC systems. Jobs are submitted via a batch system, which establishes a schedule that decides which jobs enter the system at which time to utilize certain resources. Scheduling and resource management systems are responsible for fairly allocating those resources according to the users' requests, the system capabilities, and a number of constraints and scheduling policies decided by the system owner. In contrast to commercial data centers, in the public funding and academic context of the HPC sites in this study, maximum system occupancy (\qty{90}{\percent} or even above) is targeted, so that typically oversubscription of user requests is the norm, and hardware overprovisioning the exception.  

While in the past a number of scheduling software solutions were used (e.g. Torque/Maui, PBS, LSF, etc.), today basically all German (and European) HPC centers employ Slurm~\citep{slurm}. This product has become the \textit{de-facto} scheduling solution in HPC. This has some advantages for the users, who can use almost the same job scripts to launch their jobs on different systems, but constrains all HPC sites to the limitations of this particular scheduling software, e.g. the limited capabilities of Slurm on dynamic scheduling or support of heterogeneous jobs. Furthermore, this bares the danger of a \textit{monoculture}, namely that there is only one private company controlling the development of the code and offering commercial support for this solution. A promising alternative is the Flux scheduler~\cite{flux}, which is planned to be first used at large scale with the upcoming \textit{El Capitan} system at LLNL.

The scheduling policies used today at German sites statically allocate a number of full nodes for a given time window, according to the user's request. But users tend to be conservative in their estimations, asking for longer time-windows than their jobs actually need. Therefore, when a job's execution finishes, most probably the scheduler has to drop the current schedule and create a new one. Furthermore, since free nodes were filled with smaller jobs by the \textit{backfiller}, the next large job will not be swiftly started but has to wait according to the original schedule. Overall, this will result in a disadvantage of large jobs and creates fragmentation of resources, that might require \textit{queue cleaning} actions to get an appropriate combination of nodes to allocate new large jobs, at the cost of idle resources.

Several studies have shown that a significant amount of user jobs do not fully utilize the node resources~\citep{peng2020, michelogiannakis2022, Li2023, maloney2024}. Even worse, recent systems come with very fat nodes hosting dozens of CPU cores, multiple GPUs, and other accelerators, making it more challenging for an individual user to fully employ all available devices. Node-sharing and co-scheduling of jobs with complementary profiles (e.g. compute bound with memory bound jobs, and CPU-only with GPU-centric jobs) could allow to increase the amount of jobs simultaneously running on the system, with the potential to increase system utilization and throughput. However, it also bears the risk of disturbing interference between those jobs to appear. %\esnote{REF needed here}

Energy-aware scheduling policies are applied in production already at several German sites. At JSC, idle nodes are powered down by Slurm, which can reduce the power draw of systems with idle times significantly. To give an example, the power consumption of the JUSUF system at JSC was reduced by \qty{25}{\percent} from roughly 660~MWh in 2022 to 495~MWh in 2023 by putting a rather large implementation effort for the power down feature, with the benefit that this now can be transferred to all given and future systems. In addition to the feature mentioned above, users are allowed to set the power per node depending on their needs, by reducing the operational frequency to the minimum possible without impacting application performance. Furthermore, topology-aware scheduling is currently employed at HLRS to allocate neighboring nodes to minimize network congestion. Note that HLRS uses a hypercube network topology, much more sensitive to increased latency across network hops than the tree-like interconnect topologies (e.g. dragonfly+) used by the other German HPC sites. LRZ has employed policies to encourage users to improve the energy efficiency of their codes, gathering mixed experiences.

In the German and European context the strong motivation to establish \textit{GreenHPC} has led to funding a number of R\&D projects targeting at improving scheduling mechanisms, to allow for more efficient resource sharing, dynamic allocation, job malleability, and enhanced energy-efficiency. The general trend in these projects goes towards developing a \textit{fully integrated system operations software stack}, in which the scheduler is coupled to the monitoring system and a data analytics infrastructure, bringing together and establishing correlations between job profiles, hardware capabilities, and energy use. The goal is to increase the scheduler's \textit{intelligence}, by feeding it continuously with system and data center monitoring information. Analysing extensive monitoring data will help to understand the actual resource utilization per job to adapt/optimize scheduling decisions on the fly and provides the potential to even predict the behaviour of future jobs for a more energy and resource-efficient scheduling.
The BMBF EE-HPC project~\citep{eehpc} develops a production-grade framework for energy-efficient HPC operation by continuously adopting job-specific power-cap settings through direct optimization using hardware performance counter metrics.
This framework allows to set optimal power-cap settings for every job with regard to energy-delay product, but also can be used to enforce a global system power-cap limit.

Very large HPC infrastructures, which are themselves a major power consumer, are considering using such \textit{intelligent integrated solutions} to react to the demands in the national electricity network, in a similar manner as large industrial infrastructures do. Increasing operations during phases with electricity overproduction (e.g. sunny/windy days with high electricity producing via renewable energies), would allow optimal HPC operations not only from the pure energy consumption perspective, but also reducing CO${}_2$ emissions and minimizing costs. Such advanced approaches pose difficult challenges to HPC scheduling systems and their capability to predict future system behavior, adding another layer of complexity well beyond what HPC operation tools are currently able to cope with. %\ghnote{Vielleicht sollte man noch einen Satz dazu schreiben, dass so etwas einige Challenges in Bezug auf Scheduling und Vorhersagbarkeit beinhaltet. Man kann sich viel überlegen, aber die Praxis des Rechenzentrumsbetriebs lässt vieles nicht zu. }

\section{Programming and Algorithms} % - [1000 words]}
\label{sec:prog}

With continuous improvements in HPC hardware and infrastructure, another important contributor to efficiency gains is related to the scientific workloads. By leveraging the hardware improvements effectively for better performance, shorter runtimes and thus, in most cases, more energy efficiency can be achieved. There are various approaches to achieving better hardware utilization through means of software, a fundamental component of HPC.

Modern processors with many concurrent execution threads, like many-core CPUs or massively-parallel GPUs, are enabled through effective programming models, making the distinct features of the devices abstractly available to programmers. The models are plentiful~\cite{10.1145/3624062.3624178} and include, for example, CUDA~\cite{cuda} and HIP~\cite{hip}, used for NVIDIA and AMD GPUs, respectively, which promote the \emph{SIMT} programming model to drive the many threads of the GPUs individually; or SYCL~\cite{sycl}, which is more general and provides building blocks for execution of parallel algorithms with implicit parallelization on many different hardware platforms. Through these models, programmers can effectively utilize the underlying hardware resources and gain both performance and energy efficiency improvements. 
While lower-level approaches like CUDA can serve one specific hardware design, the higher-level approaches of SYCL, Kokkos~\cite{kokkos}, or ALPAKA~\cite{alpaka} allow for more portability between hardware designs.
The convergence towards higher-level, abstract programming models eases the burden of individual programmers by relying on pre-developed primitives with excellent performance, prepared by a community of hardware-conscious engineers. At parts, the development of implementations of these programming models are conducted in the open as open source software, levering the benefits of open discussions and support by the large community~\cite{kokkos,adaptivecpp}. 
Another prominent high-level programming model is OpenMP~\cite{openmp}, which is arguably the easiest entry-point for parallel programming and enables generation of parallel code identified by the compiler.
These high-level abstractions lower the barriers to entry for new HPC users and offer productive ways for performance-conscious programming. Interoperability with low-level/native programming models can be exploited by advanced performance engineers to further optimize performance and energy efficiency.

Further abstraction with potential for even more efficiency gains can be achieved by increasingly relying on libraries. These packages, sometimes only available in binary form from hardware vendors, offer implementations of algorithms and patterns, usually with high grade of sophistication and deep hardware focus. A famous example are the BLAS APIs, which are implemented for nearly all hardware devices by their vendors, e.g., oneMKL (Intel)~\cite{mkl} and cuBLAS (NVIDIA)~\cite{cublas}. Of course, also here, community-driven libraries in the open-source realm exist, like BLIS~\cite{blis}, an extensible library with many hardware-specific micro-kernels~\cite{nassyr2023programmatically,nassyreuropar}. Libraries exist beyond computational algorithms. NCCL~\cite{nccl}, %\esnote{what about RCCL?}\ahnote{RCCL is a 'simple' fork of NCCL, and there's also oneCCL; I think NCCL is best-known here and should be sufficient as an example}
 for example, can improve GPU-centric communication significantly over the HPC-default MPI, in AI applications and beyond~\cite{chase}.

Separation of performance-critical code and user-facing interfaces offers a productive way to gain performance, ideally relying on well-defined APIs implemented by advanced libraries. The approach enables moving away from strict, compiled languages like C++, towards more agile languages with shorter turn-around times, like Python. The same approach essentially enables the ongoing AI revolution, in which frameworks such as PyTorch implement the user-facing API, using hardware-specific libraries (like MlOpen~\cite{mlopen} and oneDNN~\cite{onednn}) in the back-end, augmented with just-in-time (JIT) compilation facilities for customization.
Julia is another high-level programming language which is JIT-compiled, offering a productive alternative to the classical low-level languages~\cite{hunold2020benchmarking,juliateich}. %\esnote{Should we say something about the energy efficiency of python? I have heard often criticism about it...}\ahnote{I tried writing something, but it make the text lenghty. Isn't this said already at the beginning of the paragraph regarding the separation of performance critical code and interface?}

Many of the hardware-centric advancements are in code generation.
Some domain-specific frameworks exist to generate low-level, hardware-optimized code through higher level definitions. Examples are \emph{gt4py}~\cite{gt4py} for the climate domain or \emph{pystencils}~\cite{pystencils} for, among others, numerical CFD simulations; both optimize operations on a grid. Even generic domain-specific languages exist, like \emph{AnyDSL}~\cite{anydsl}, which utilizes partial evaluation to generate hardware-specific code.
All frameworks, runtimes, and (JIT)~compiled programs benefit from continuously maturing compiler infrastructures, especially open source software. Of distinct importance is LLVM~\cite{llvm}, which has emerged as the key enabler of modern compilation workflows, covering language-specific frontends (e.g. C++, Fortran), advanced optimizations in the intermediate languages (e.g. MLIR), versatile back-ends for many hardware architectures (e.g. x86, RISC-V, PTX), linking (e.g. LTO), and runtimes (like OpenMP).
The open-source-nature of the compiler infrastructure allows contributions from research institutions, industrial corporations, and hardware vendors.

In the heterogeneous, quickly evolving hardware landscape, co-design is essential to align application requirements and hardware opportunities.
Close collaboration between hardware consumers and producers allows informing hardware-fitting algorithms, enablement through effective compiler and framework/library support, and design decisions for the hardware. An example of a hardware-conscious implementation is temporal blocking of sparse matrix power kernels, which enables cache reuse and corresponding performance gains for traditionally memory-bound algorithms~\cite{Alappat2023,Alappat2024}. 

A possible path to understanding application performance on specific hardware is thorough analytic, white-box performance models. Such models abstract the intricacies of the hardware, the software, and their interactions. A prominent example is the \emph{Roof{}line Model}, which only assumes two hardware bottlenecks (computational peak performance and memory data transfer) and reduces the software to a single number (operational intensity)~\cite{williams2009}. Many extensions of the Roof{}line Model exist and can be coupled with energy models~\cite{Hager2012,Hofmann-18-1}. Although such white-box models lack accuracy due to their inherent simplifications, they are indispensable for guiding design decisions for applications without time-consuming and complex in-depth research.  A key component of this process is \emph{bottleneck awareness}, which not only allows to select the most appropriate hardware but also the energy-relevant execution modalities such as clock frequency (DVFS) and number of cores per chip (concurrency throttling)~\cite{Hager2012, Wittmann2016}. 

Selecting hardware components and a system design appropriate for the site's user portfolio is also crucial to maximize energy efficiency in real-world operations. Therefore, in addition to synthetic benchmarks (e.g. High Performance Linpack (HPL) or HPCG), application-based benchmarks are typically included in the procurement and acceptance of HPC systems~\cite{herten2024} as well. The selection of use cases in the procurement benchmark suite represents the current major consumers of computing time, but also the expected evolution of the user portfolio, e.g. a larger set of deep learning training applications.  
Different components or phases of a single application may have unique hardware requirements. For example, the ICON climate and weather model offers opportunities for substantial energy efficiency gains when running on heterogeneous hardware. This is because the model's coarse-grained task parallelism, comprising different components such as atmosphere, ocean, radiation, and biogeochemical modules, allows for tailored deployment on the most suitable hardware, reducing overall energy consumption. However, a major challenge is to support the various programming models that are required for this optimisation~\cite{gmd-2024-54}. This task is being tackled as part of the \textit{GreenHPC} project EECliPs, for example.

The techniques for increasing performance and energy efficiency on modern hardware devices are quite involved and complex. It requires a well-educated workforce. Starting at university, where HPC-related courses are offered, but continuing during the professional career, where HPC basics and advanced topics are covered in courses. The German HPC community, especially the Gauss Centre for Supercomputing (GCS) and the NHR Alliance of Tier-2 centers but also the local HPC networks of the federal states, offer a plethora of courses for HPC users and developers on various skill levels. The centers work with the HPC Certification Forum to map these courses to a tree of well-defined skills, enabling the attendees to compile the course program that fits their needs best. %\ghnote{Do we need links here?}\ahnote{Not sure, possibly not? Or maybe a link to the EuroCC list? It should incorporate GCS and NHR courses, AFAIK: \url{https://www.eurocc-access.eu/services/training/}}

\section{Conclusions and future work} % - 800 words}
\label{sec:concl}

The number and size of HPC infrastructures are growing, driven by the demand for compute time in the HPC and AI communities. This trend, together with increasing environmental concerns, electricity prices, and regulatory policies, is driving HPC hosting sites to develop and implement strategies to maximise the energy efficiency of their operations. The experience of German centres hosting jointly $\approx$\qty{300}{\peta\flop\per\second} compute performance (status 2024) shows that no single measure alone can achieve the efficiency targets set. A holistic approach to energy efficiency is required, including green power supply, power capping, optimised cooling systems, heat recovery, careful selection and combination of processing and storage technologies, comprehensive monitoring software, advanced scheduling and resource management techniques, and application optimisation measures. 

The German sites reporting here are already implementing solutions in all these areas and have concrete plans for future extension. Direct liquid cooling strategies are already in place and waste heat is reused at many sites. Hardware heterogeneity, especially in terms of processing units, is increasing as the end of Moore's Law makes higher efficiency per component possible only through specialisation. Heterogeneous system architectures are widely used, where processing heterogeneity is organised in partitions or compute modules. Maximising resource utilisation in such systems requires advanced scheduling and resource management techniques, with increasing interest on node sharing and dynamic scheduling mechanisms. Comprehensive monitoring solutions for data centre, system components and user jobs are utilized; however, homogenisation and closer cooperation in this area could be beneficial. Last but not least, the HPC sites are investing significant efforts in user support to optimise application codes for highest performance, and are also developing optimised libraries, programming models and tools to improve energy efficiency from the user perspective. 

Looking into the future, the HPC sites are involved in a number of research projects and initiatives to further improve energy efficiency in HPC. Data centre upgrades include medium- and long-term plans to increase the rate of waste heat reuse, but these plans could be jeopardised by the increasing power density of newer processors (especially GPUs) and their need for lower cooling temperatures. The choice of processing technologies for future system upgrades will largely be based on the highest achievable performance per Watt. As a result, future systems are likely to be more heterogeneous in terms of hardware, and there will be a need for more dynamic resource allocation, possibly taking into account variations in the power grid. Such load management and advanced scheduling software does not yet exist, or at least not in production quality. Therefore, a number of research and development projects are underway in Germany and Europe-wide to develop integrated system monitoring and management solutions to address this need. Other projects are devoted to programming models, algorithms and tools for improved application efficiency, since these require ongoing efforts to adapt to new hardware features. The use of arithmetic with reduced precision has attracted particular interest, which is supported by the growing trend towards ML applications and specialised hardware.

Energy efficiency will remain a key issue in the design, deployment, operation and use of future HPC systems, requiring the attention of all stakeholders: component suppliers, system integrators, hosting site operators, system software designers, programming model engineers, application developers, and users.

\section*{Conflict of Interest Statement}

The authors declare that the research was conducted in the absence of any commercial or financial relationships that could be construed as a potential conflict of interest.

\section*{Author Contributions}

This paper is the result of a collaborative effort between eight HPC sites in Germany. Data in tables~\ref{tab:enbreakdown}, \ref{tab:DLC}, \ref{tab:HeatReuse} and \ref{tab:monitor} have been collected with input from all sites. Further individual contributions from the authors are: ES has developed the concept, coordinated the overall work, and reviewed the content, she is also main author of sections~\ref{sec:intro}, \ref{sec:syshw}, \ref{sec:sched} and \ref{sec:concl} and contributor to section~\ref{sec:motiv}; HB and PG have contributed data and technological insights and application (climate related) knowledge from DKRZ. NE has contributed to section~\ref{sec:sched} and reviewed the paper; JE is the author of figures~\ref{fig:EnergyEfficiencyChips} and \ref{fig:monitoring-architecture} and has contributed to section~\ref{sec:monit}; SES and CM are the main authors of section~\ref{sec:stor}, to which SO has also contributed; TF, PF and BvSV have contributed with JSC data in section~\ref{sec:infra} and reviewed the overall content; MF has contributed data from KIT and is main author of section~\ref{sec:motiv}; DH is main author of sections~\ref{sec:cool} and \ref{sec:reuse}; GH has contributed to section~\ref{sec:prog}; AH is main author of section~\ref{sec:prog} and has contributed to section~\ref{sec:motiv}; TI is main author of section~\ref{sec:dataagg} and has contributed to sections~\ref{sec:powmgt} and \ref{sec:monit}; BK and RS have provided HLRS data and insights, and reviewed the paper; EL and KR have provided MPCDF data and information and contributed particularly to sections~\ref{sec:infra} and \ref{sec:monit}; MO  is main author of section~\ref{sec:monit} and has contributed to sections~\ref{sec:infra} and~\ref{sec:powmgt}; and KT has contributed to section~\ref{sec:monit}.

\section*{Funding}

The results of the DKRZ are part of the work from the \textit{GreenHPC} projects EE-HPC (grant number 16ME0586) and EECliPs (grant number 16ME0599K), funded by the BMBF. ClusterCockpit development is funded as part of the \textit{GreenHPC} project EE-HPC (grant number 16ME0586).

\section*{Acknowledgments}

\section*{Supplemental Data}

\section*{Data Availability Statement}
The datasets for figures~\ref{fig:EnergyEfficiencyChips} and ~\ref{fig:electriticy} can be provided on a public repository, upon paper acceptance.

\bibliographystyle{Frontiers-Harvard} %  Many Frontiers journals use the Harvard referencing system (Author-date), to find the style and resources for the journal you are submitting to: https://zendesk.frontiersin.org/hc/en-us/articles/360017860337-Frontiers-Reference-Styles-by-Journal. For Humanities and Social Sciences articles please include page numbers in the in-text citations 
\bibliography{ref}

\end{document}